\documentclass[pre,aps,twocolumn,showpacs,floatfix,10pt,superscriptaddress]{revtex4-1}
\usepackage[dvips]{graphicx}
\usepackage{amsmath}
\usepackage{amssymb}
\usepackage{color}

\begin{document}

\title{Continuous time random walks under power-law resetting}
\author{Anna S. Bodrova}
\affiliation{Humboldt University, Department of Physics, Newtonstrasse 15, 12489 Berlin, Germany}
\author{Igor M. Sokolov}
\affiliation{Humboldt University, Department of Physics, Newtonstrasse 15, 12489 Berlin, Germany}
\affiliation{IRIS Adlershof, Zum Gro{\ss}en Windkanal 6, 12489 Berlin, Germany}

\begin{abstract}
We study continuous time random walks (CTRW) with power law distribution of waiting times under resetting which 
brings the walker back to the origin, with a power-law distribution of times between the resetting events. Two situations are considered. 
Under complete resetting, the CTRW after the resetting event starts anew, with a new waiting time, independent of the prehistory.
Under incomplete resetting, the resetting of the coordinate does not influence the waiting time until the next jump.
We focus on the behavior of the mean squared displacement (MSD) of the walker from its initial position, on the conditions 
under which the probability density functions of the walker's displacement show universal behavior, and on this universal behavior itself. 
We show, that the behavior of the MSD is the same as in the scaled Brownian motion (SBM), being the mean field model of the CTRW.
The intermediate asymptotics of the probability density functions (PDF) for CTRW under complete resetting (provided they exist) are also the same as in the
corresponding case for SBM. For incomplete resetting, however, the behavior of the PDF for CTRW and SBM is vastly different.
\end{abstract}
\maketitle

\section{Introduction}

Recently, there is a splash of interest in statistical properties of different stochastic processes
under resetting, when a random process is interrupted by 
a resetting event, and restarts anew from prescribed initial conditions. 
The interest to this kind of processes is nurtured by their abundance in nature and by their importance in search, see \cite{review} for the review. 
The situation is mostly exemplified by a time-dependent position of a particle which performs some kind of random motion
and returns to the origin on the resetting event. 
The random motion under stochastic resetting can thus be considered as the interplay of two distinct random processes: 
the resetting process, a point process on the real line representing the time axis, and particle's motion between the resetting events, the displacement process. 

The waiting time distribution function between two resetting events can
be exponential \cite{EvansMajumdar}, deterministic (the most effective one for the search processes) \cite{shlomi2017}, power-law \cite{NagarGupta} or of
other type \cite{shlomi2017,palrt,res2016,shlomi2016}. Most studies treat the resetting as an instantaneous event \cite{review}, but also the situations when 
some time is needed by the particle to come back to the initial position were considered \cite{me,me2,shlomi, shlomi1,shlomi2,campos}. The first study of resetting has been 
devoted to Brownian motion \cite{EvansMajumdar} as a displacement process, later the discussion has been generalized for other types of motion, 
such as L\'evy flights \cite{levy1,levy2}, L\'evy walks \cite{china}, scaled Brownian motion (SBM) \cite{AnnaNonrenewal, AnnaRenewal}, and
continuous time random walks (CTRW) \cite{MV2013,MC2016,Sh2017,ctrw,ctrwres}. This last situation is the topic of the present work. 

Continuous time random walk (CTRW) is a process, when the time of the next step of a random walk is chosen according to a certain probability distribution \cite{sokbook, MontrollWeiss}. The
applications of CTRW range from charge carrier motion in disordered semiconductors \cite{disordered} to earthquake modeling \cite{earthquake1, earthquake2}, biology \cite{bio}, and economics
\cite{eco1,eco2}. The properties of CTRW with an exponential waiting time density $\psi(t)=re^{-rt}$ correspond to normal diffusion \cite{sokbook}, with the mean squared displacement
(MSD) $\langle x^2 (t) \rangle$ growing linearly in time, but the properties of
CTRW with a power-law waiting time probability density function (PDF) $\psi (t) \sim t^{-1-\alpha}$ (with $0< \alpha < 1$) are quite different, giving rise to a slower, subdiffusive behavior with 
$\langle x^2 (t) \rangle \propto t^\alpha$. The properties of such subdiffusive CTRW under Poissonian resetting were recently considered in Ref. \cite{ctrwres}, 
providing a nice introduction to the problem of resetting in CTRW.
 
On the mean field level some properties of CTRW (for example, its aging) resemble those of  subdiffusive scaled Brownian motion (SBM), a diffusion process with the time-dependent diffusion coefficient $D(t)\sim
t^{\alpha-1}$ and the mean-squared displacement (MSD) $\left\langle x^2(t)\right\rangle\sim t^{\alpha}$ \cite{SokolovSBM}.  The SBM is a Markovian process, while CTRW is a non-Markovian (semi-Markovian) one.
Both random processes, the CTRW and the SBM are processes with non-stationary increments. However, in SBM this non-stationarity is modeled via the explicit time dependence of the diffusion coefficient, 
while the CTRW, being of the renewal class, lacks explicit time dependence of its parameters. Therefore, some properties of the processes (for example, their behavior under
confinement) differ \cite{MetzlerSBM}. SBM can be used in order to describe the dynamics of granular gases \cite{annapccp,annagg,hadiseh,ultraslow}.

The non-stationarity of increments of the displacement process leads to two different situations under resetting, which were indistinguishable if the increments of the displacement process were stationary. 
The first one corresponds to the case when the memory on the course of the displacement process preceding the resetting event is fully erased, and the second one to the case when this 
memory is partially retained: The dynamics of the underlying process can be either rejuvenated after resetting or not influenced by the resetting of the coordinate. We will 
refer to the first case as to the one of \textit{complete resetting}, while the second case will be referred to as the case of \textit{incomplete resetting}.
In SBM these two situations correspond to the cases when the time dependent diffusion coefficient $D(t)$ also resets to the initial value $D(0)$ together with the coordinate
of the particle \cite{AnnaRenewal}, or remains unaffected by the resetting events \cite{AnnaNonrenewal}. In CTRW the first assumption corresponds to a situation 
when the resetting interrupts the waiting period between the jumps, and, after the resetting, a new waiting time is chosen independently from the prehistory of the process.  
The second assumption implies that the waiting period started before the resetting event is not interrupted by the resetting. 
These two cases has been investigated and compared for the CTRW with exponential resetting \cite{ctrwres}. In the current study we investigate the behavior of subdiffusive CTRW 
under resetting process with power-law distribution of times between the resetting events, compare the results with such for SBM, and discuss similarities and differences between 
the behavior of CTRW and its mean field model. We show that the behavior of the MSD in both processes is similar. Considerable similarities are also found in the intermediate 
asymptotic behavior of the probability density functions (PDFs) of both processes under complete resetting, while for incomplete resetting CTRW shows additional fluctuation 
effects leading to strong differences in PDFs. 

The further structure of the work is as follows: In Sec. \ref{sec:Gen} we define the models and introduce the notation. The behavior of the MSD is then discussed in Sec. \ref{sec:MSD}.
The properties of the intermediate asymptotics of the PDFs and the conditions under which these can be observed are discussed in Sec. \ref{sec:PDFs}. The conclusions follow in Sec. \ref{sec:Concl}.

\section{Model}
\label{sec:Gen}

\subsection{Continuous time random walks}

A standard (``wait-first'') CTRW starts at $x=0$ at time $t_0$ (in a situation without resetting this is typically put to zero) with the waiting time
\cite{ScherMont}. Other variants of the CTRW include the walks starting from a jump (similar to the corresponding correlated model of \cite{Magdziarz}), the walks anticipating the
next jump after
the observation time $t$ (``oracle'' walk) and other clustered models, \cite{Jurlewitz}. The CTRW by itself may be considered as an interplay (subordination) of two
distinct random processes: The \textit{parent process}, being a simple random walk with discrete steps, and the \textit{directing process} (subordinator, operational time) defining the 
random number of steps the parent process made up to the physical time $t$. In this work we will consider resetting of the classical Scher-Montoll wait-first scheme,
although the jump-first variant will appear at intermediate steps of our discussion.

Although general expressions may be obtained in a Fourier-Laplace domain, like it was done in Ref. \cite{ctrwres}, these,
for the case when the resetting times follow a power-law distribution, are difficult to analyze. Therefore, for getting asymptotic expressions for PDFs we will use the real space / time domain approach, relying on the asymptotic form of the CTRW's PDFs.
Therefore, the methods applied in the present work differ considerably from those used previously.  

We study power-law distribution of the CTRW steps:
\begin{equation}
 \psi(t) = \frac{\alpha t_0^\alpha}{(t_0 + t)^{1+\alpha}}\,.
 \label{eq:CTRWpar}
\end{equation}
Here $t_0$ is the characteristic time of the power-law decay connected with the median
value $m_t$ of the waiting time via $m=(2^{1/\alpha}-1) t_0$.
The survival probability $\Psi(t)$ gives the probability that no stepping occurs between 0 and $t$:
\begin{equation}
\Psi (t) = 1 - \int\limits_0^t {\psi (t')dt'}  = \int\limits_t^\infty  {\psi (t')dt'}\,.
\end{equation}
For the power-law distribution of the waiting times it scales also acccording to the power law:
\begin{equation}
\Psi(t) = \frac{t_0^\alpha}{(t_0 + t)^\alpha}\,.
\end{equation}
It is convenient to to switch between the time and the Laplace domains. The Laplace transform of the resetting PDF is
\begin{equation}
\tilde{\psi}(s)=\int_0^{\infty}\psi(t)\exp(-ts)dt
\end{equation}
and the Laplace transform of the survival probability can be expressed via $\tilde{\psi} (s)$ as
\begin{equation}
\tilde{\Psi} (s) = \frac{1-\tilde{\psi}(s)}{s}\,.
\end{equation}
For $\alpha < 1$ the asymptotics of the Laplace transform of $\Psi(t)$ is $\Psi(s) \simeq \Gamma(1-\alpha) s^{\alpha -1} t_0^\alpha$, and $\psi(s) \simeq 1 -
\Gamma(1-\alpha) s^\alpha t_0^\alpha$.
The probability density $\psi_n(t)$ that the $n$-th resetting event happens at time $t$ satisfies the renewal equation \cite{sokbook}
\begin{equation}
\psi_n(t) = \int_0^{t}\psi_{n-1}(t^{\prime})\psi(t-t^{\prime})dt^{\prime}\,,
\end{equation}
and the sum of all $\psi_n(t)$ gives the rate of resetting events at time $t$: 
\begin{equation}\label{rate}
\mu(t)=\sum_{n=1}^{\infty}\psi_n(t)\,.
\end{equation}
Its Laplace transform yields
\begin{equation}
\tilde \mu (s) = \sum\limits_{n = 1}^\infty  {{{\tilde \psi }^n}(s)}  = \frac{{\tilde \psi (s)}}{{1 - \tilde \psi (s)}}\,.
\end{equation}
The stepping rate for $\alpha < 1$ is given by
\begin{eqnarray}
 \mu(s) &\simeq& \frac{1}{\Gamma(1-\alpha)} (s t_0)^{-\alpha} \nonumber  \\ 
 \mu(t) &\simeq& \frac{\sin \pi \alpha}{\pi} t_0^{-\alpha} t^{\alpha -1}
 \label{eq:mu}
\end{eqnarray}
in the Laplace domain and in the time domain, respectively. 

For $\alpha > 1$ (the case which would correspond to normal diffusion for CTRW without resetting) we have
\begin{equation}
 \mu(t) = \frac{\beta - 1}{\tau_0}
\end{equation}

\subsection{Complete and incomplete resetting}

As we already mentioned, two situations are considered. In the first one, after a resetting the CTRW process
starts anew, from a new waiting time which is independent of the prehistory of the process (complete resetting). This case corresponds to the first model of Ref. \cite{ctrwres} 
and will be denoted as case (1) in the text and in figures.
The case of incomplete resetting (case (2)) corresponds to the second model of Ref. \cite{ctrwres}. In this case the coordinate of the walker is set to zero under resetting, which however does not interrupt the waiting period. In this case the memory on the beginning of the waiting period of the CTRW is not erased. 

\begin{figure}[tbp]
\begin{center}
\scalebox{0.4}{\includegraphics{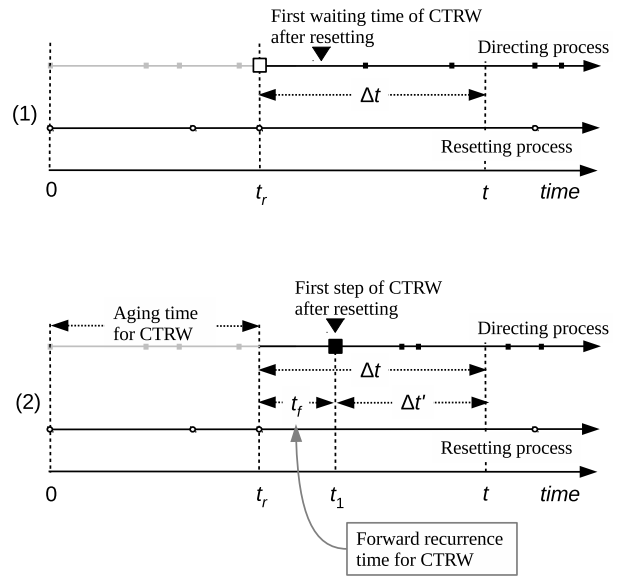} }
\caption{The event diagrams of CTRW under complete (1) and under incomplete (2) resetting. The renewal events of the CTRW are denoted by black and gray filled squares, the 
renewals of the resetting process are denoted by empty circles. The difference between the two situations
is that the time of the last renewal of the resetting process before the observation is the time of the (new) beginning of the
wait-first CTRW in (1), and the aging time of the CTRW in (2). The new beginning of the CTRW in (1) is denoted by a larger empty square (there is no jump taken at this time),
and the time $\Delta t = t-t_r$ corresponds to the observed duration of the wait-first CTRW.
In (2) the time of the last resetting is the aging time for the CTRW, and the first jump of the CTRW takes place at time $t_1$, denoted by a 
larger black square.The observed duration of the jump-first CTRW $\Delta t'$ corresponds to $t -
t_1$. \label{fig:Events}}
\end{center}
\end{figure}

The event diagrams, showing the temporal order of jumps of CTRW and resetting events for the two models are displayed in Fig. \ref{fig:Events} and elucidate the 
notation used. Thus, in the case (1) a wait-first (standard) CTRW starts anew at time of the last resetting event $t_r$.
The total duration of the observed part of the CTRW (which is the time interval between the last resetting event at $t_r$ and the time $t$ at which the 
position of the walker is measured) is equal to $\Delta t = t-t_r$. 
For the case (2) of incomplete resetting, the resetting time falls into a waiting time between the two steps of the CTRW (or in the 
very first waiting time between the preparation and the first step), which is not interrupted by the resetting event. 
 In this case we consider a jump-first CTRW starting at this forward recurrence time of a CTRW following the time of the last resetting. 
The total duration of the observed part of this jump-first CTRW is then $\Delta t' = t - t_r - t_f$. Since the time  $t_r$ of the last resetting event now corresponds to the aging time of the CTRW, the waiting time for the first step in CTRW after resetting will typically be longer than in the previous case due to aging effects \cite{sokbook}, provided the second moment of the waiting time is large enough or diverges.

\subsection{Power-law resetting}

The waiting time PDF of the resetting process will be denoted by $\phi(t)$ and is distributed according to the power law distribution function
\begin{equation}
 \phi(t) = \frac{\beta \tau_0^\beta}{(\tau_0 + t)^{1+\beta}}\,.\label{eq:respar}
\end{equation}
Here $\tau_0$ is the characteristic time of the power-law decay connected with the median
value $m_{\tau}$ of the resetting time via $m_{\tau}=(2^{1/\beta}-1)\tau_0$.
The survival probability $\Phi(t)$ gives the probability that no resetting event occurs between 0 and $t$,
\begin{equation}
\Phi (t) =  \int\limits_t^\infty  {\phi (t')dt'}\,.
\end{equation}
For the power law distribution of waiting times it also scales according to the power law:
\begin{equation}\label{survres}
\Phi(t) = \frac{\tau_0^\beta}{(\tau_0 + t)^\beta} 
\end{equation}
For the case of the power-law PDF it is convenient to to switch between the time and the Laplace domains.
The rate of resetting events at time $t$ may be obtained analogously to the stepping rate of the CTRW. For $\beta < 1$ the resetting rate is time-dependent and is given by
\begin{eqnarray}
 \kappa(s) &\simeq& \frac{1}{\Gamma(1-\beta)} (s \tau_0)^{-\beta} \nonumber  \\ 
 \kappa(t) &\simeq& \frac{\sin \pi \beta}{\pi} \tau_0^{-\beta} t^{\beta -1}
 \label{kappab0}
\end{eqnarray}
in the Laplace domain and in the time domain, respectively. 

For $\beta > 1$ the rate of resetting events stagnates for long $t$ and is given by 
\begin{equation} \label{kappab1}
 \kappa(t) = \frac{1}{\langle t \rangle} = \frac{\beta - 1}{\tau_0}
\end{equation}
with $\langle t \rangle$ being the mean waiting time between two resetting events. 



\section{Mean number of steps and the MSD}
\label{sec:MSD}


\subsection{MSD of free CTRW}
The MSD in a free CTRW is proportional to the mean number of steps \cite{sokbook}
\begin{equation}
 \langle x^2(t) \rangle = a^2 \langle n(t) \rangle
 \label{eq:MSDCTRW}
\end{equation}
where $a^2$ is the mean squared displacement in a single step.
The mean number of steps performed up to time $t$ can be obtained as the integral of the stepping rate (Eq.~\ref{rate}):
\begin{equation}
 \langle n(t) \rangle = \int_0^t \mu(t') dt'
 \label{eq:meann}
\end{equation}
For $\alpha < 1$ it is equal to
\begin{equation}\label{nal1}
 \langle n(t) \rangle = \frac{\sin \pi \alpha}{\pi \alpha} \left(\frac{t}{t_0} \right)^\alpha
\end{equation}
and for $\alpha > 1$
\begin{equation}\label{nag1}
 \langle n(t) \rangle \simeq (\alpha -1)\frac{t}{t_0}
\end{equation}
Both expressions hold for $ t \gg t_0$.

The coefficient of anomalous diffusion $K_\alpha$ is normally defined via
\begin{equation}
  \langle x^2(t) \rangle = 2 K_\alpha t^\alpha,
\end{equation}
so that for $\alpha<1$
\begin{equation}
 K_\alpha = \frac{1}{2} \frac{\sin \pi \alpha}{\pi \alpha} \frac{a^2 }{t_0^\alpha}.   \label{eq:DiffKoeff}
\end{equation}
and for $\alpha>1$  the (normal) diffusion coefficient reads
\begin{equation}
 K_\alpha = K_1 =  \frac{\alpha -1}{2} \frac{a^2}{t_0}.
\end{equation}

\subsection{MSD of CTRW with resetting}

For a non-biased CTRW with resetting the mean squared displacement, both for aged, and for non-aged situation is proportional to the mean number of steps made during the observation time \cite{SoBluKla},
\begin{equation}\label{xan}
 \langle x^2 (t) \rangle = a^2 \langle \langle n_{1,2}( \Delta t ) \rangle \rangle,
\end{equation}
where $a^2$ is the mean squared length of the step, and the indices $1$ and $2$ denote the complete and incomplete resetting (See Fig.~\ref{fig:Events}). The double average on the right-hand site of Eq.~(\ref{xan}) is taken over the realizations of the directing process of the CTRW (i.e. over the CTRW waiting times), and
over the duration
$\Delta t$ of the period between the last resetting and the observation time. 
For given $t_r$ (or $\Delta t$),
the mean numbers of steps $n_1 (\Delta t)$ for the complete resetting (the average over all possible realization of the waiting times of the directing process of the CTRW) 
is given by
\begin{equation}
 \langle n_1 (\Delta t) \rangle = \langle n(\Delta t) \rangle
\end{equation}
with $\langle n(\Delta t) \rangle$ given by Eq.(\ref{eq:meann}). For the incomplete resetting we have for the same single average (See Fig.~\ref{fig:Events}, panel (2))
\begin{equation}
  \langle n_2 (\Delta t) \rangle = \langle n(t) \rangle - \langle n(t_r) \rangle. 
  \label{eq:MMDeltat}
\end{equation}
The double means we are interested in are obtained by averaging these means over the distribution of $\Delta t$ or $t_r$. For the first case of complete resetting we obtain
\begin{equation}\label{n1}
 \langle \langle n_1( \Delta t ) \rangle \rangle = \int_0^t \langle n( \Delta t ) \rangle p_1(\Delta t |t) d \Delta t
\end{equation}
Here $\langle n_1( \Delta t ) \rangle$ is given by Eq.~(\ref{nal1}) for $\alpha<1$ and by Eq.~(\ref{nag1}) for $\alpha>1$.
For the second case of incomplete resetting we get
\begin{equation}\label{n2}
 \langle \langle n_2( \Delta t ) \rangle \rangle = \langle n( t ) \rangle- \int_0^t \langle n(t_r) \rangle  p_2(t_r|t) d t_r
\end{equation}
The PDF $ p_2(t_r|t)$ of the last resetting before the observation at time $t$ is given by
\begin{equation}\label{p2initial}
 p_2(t_r|t) = \kappa(t_r) \Phi(t-t_r).
\end{equation}
The meaning of this equation is that the $\kappa(t_r)dt_t$ defines the probability to have a resetting event between $t_r$ and $t_r + dt_r$, and $\Phi(t-t_r)$
the survival probability that no resetting event took place afterwards (Eq.~\ref{survres}).
The distribution of the duration of the part of CTRW observed after the resetting $\Delta t = t-t_r$ follows by the change of variables:
\begin{equation}\label{p1initial}
 p_1(\Delta t|t) = \kappa(t-\Delta t) \Phi(\Delta t).
\end{equation}
The information about mean number of steps will be also important for the next section \ref{sec:PDFs}. The calculation of the probability distribution functions is performed under assumption that this number of steps is large. Only under this condition the universal (not dependent on microscopic parameters) intermediate asymptotics can appear.

\subsection{Mean number of steps for $0 < \beta < 1$ } 

The distribution of $\Delta t$ at given $t$ for complete resetting (case 1) is given by inserting Eq.~(\ref{kappab0}) and Eq.~(\ref{survres}) into Eq.~(\ref{p1initial}) and for longer $\Delta t$ gets independent from $\tau_0$:
\begin{equation}\label{p1b0}
 p_1(\Delta t | t) \simeq \frac{\sin \pi \beta}{\pi} (t-\Delta t)^{\beta -1} \Delta t^{-\beta}. 
\end{equation}
For the case 2 of the incomplete resetting the distribution of the aging time $t_r$ for given $t$ is given by the similar expression
\begin{equation}\label{p2b0}
 p_2(t_r | t) \simeq \frac{\sin \pi \beta}{\pi}  (t_r)^{\beta -1} (t-t_r)^{-\beta}.
\end{equation}

\subsubsection{Subdiffusion with $0 < \alpha <1$ } 
For the subdiffusive CTRW with \textbf{complete resetting} we insert Eq.~(\ref{nal1}) and Eq.~(\ref{p1b0}) into Eq.~(\ref{n1}) and obtain after straightforward algebra
\begin{equation}
\langle \langle n_{1}( \Delta t ) \rangle \rangle = \left(\frac{t}{t_0} \right)^\alpha \frac{\sin \pi \alpha}{\pi \alpha}  \frac{\sin \pi \beta}{\pi} \mathrm{B}(\beta, 1+\alpha - \beta).
\end{equation}
For the subdiffusive CTRW with \textbf{incomplete resetting} we introduce Eq.~(\ref{nal1}) and Eq.~(\ref{p2b0}) into Eq.~(\ref{n2}) and get
\begin{equation}
\langle \langle n_{2}(t) \rangle \rangle =  \frac{\sin \pi \alpha}{\pi \alpha}  \left( \frac{t}{t_0} \right)^\alpha \left[1 - \frac{\sin \pi \beta}{\pi} \mathrm{B}(\alpha+ \beta,1-\beta)  \right].
\end{equation}
The fact that resetting with $0 < \beta < 1$ does not change the power-law behavior in MSD is analog to
the observation for SBM. \\

\subsubsection{Normal diffusion with $\alpha \geq 1$ } 

Let us consider at first the case of the \textbf{complete resetting}. Inserting Eq.~(\ref{nag1}) and Eq.~(\ref{p1b0}) into Eq.~(\ref{n1}) we get
\begin{equation}\label{n1nice}
 \langle \langle n_{1}( \Delta t ) \rangle \rangle = \frac{\alpha - 1}{t_0} \frac{\sin \pi \beta}{\pi}\int_0^t \Delta t^{1 - \beta}  (t- \Delta t)^{\beta -1} d \Delta t.
\end{equation}
Changing the variable of integration to $\xi = \Delta t /t$ we obtain
\begin{equation}
 \langle\langle n_{1}( \Delta t ) \rangle\rangle =\frac{\alpha - 1}{t_0} \frac{\sin \pi \beta}{\pi} t \mathrm{B}(\beta, 2 - \beta),
\end{equation}
where the beta function
\begin{equation}
 \mathrm{B}(\beta, 2 - \beta) = \int_0^1 \xi^{\beta - 1}(1-\xi)^{1-\beta} d \xi
\end{equation}
is equal to
\begin{equation}
 \mathrm{B}(\beta, 2 - \beta) = \frac{\pi \beta}{\sin \pi \beta}\,.
\end{equation}
In such a way we get
\begin{equation}
  \langle \langle n_{1}( \Delta t ) \rangle \rangle = \beta (\alpha-1)\frac{t}{t_0}.
\end{equation}
For the \textbf{incomplete resetting} substitution of Eq.~(\ref{nag1}) and Eq.~(\ref{p2b0}) into Eq.~(\ref{n2}) surprisingly leads to the same result
\begin{equation}
  \langle \langle n_{2}( \Delta t ) \rangle \rangle =\langle \langle n_{1}( \Delta t ) \rangle \rangle = \beta (\alpha-1)\frac{t}{t_0}.
\end{equation}

In both cases the mean number of steps grows with observation time, so that for these cases the intermediate asymptotics in $x$ discussed in the next section indeed appear at long times.

\subsection{Mean number of steps for \boldmath $\beta > 1$ \unboldmath .} In this case the rate of resetting events is time-independent, so that 
\begin{equation}\label{p1b1}
 p_1(\Delta t|t) = \frac{\beta -1}{\tau_0^{1-\beta}} \frac{1}{(\tau_0 + \Delta t)^\beta} 
\end{equation}
and 
\begin{equation}\label{p2b1}
 p_2(t_r | t) =  \frac{\beta -1}{\tau_0^{1-\beta}} \frac{1}{[\tau_0 + t - t_r]^\beta}
\end{equation}

\subsubsection{Subdiffusion with $0 < \alpha <1$ }  

In case of \textbf{complete resetting} we introduce Eq.~(\ref{nal1}) and Eq.~(\ref{p1b1}) into Eq.~(\ref{n1}) and get
\begin{equation}
\langle \langle n_{1}( \Delta t ) \rangle \rangle = t_0^{-\alpha} \frac{\sin \pi \alpha}{\pi \alpha} \frac{\beta -1}{\tau_0^{\beta -1}} t^{\alpha - \beta + 1} I_1(\alpha,\beta; z),\end{equation}
where $\xi = \Delta t / t$, $z = \tau_0 / t$ and the integral
\begin{equation}\label{I1}
I_1(\alpha,\beta;z) = \int_0^1 \xi^\alpha (z + \xi)^{-\beta} d \xi 
\end{equation}
will repeatedly appear in our calculations, and its asymptotic behavior
in different domains of parameters is discussed in Appendix A. 
For $\alpha > \beta - 1$ the function $I_1(\alpha,\beta;z)$ tends to a constant (see Eq.(\ref{eq:AB1}) in Appendix A) and at large
$t$ we have
\begin{equation}
\langle \langle n_{1}( \Delta t ) \rangle \rangle \simeq t_0^{-\alpha} \tau_0^{\beta-1} \frac{\beta - 1}{2 + \alpha - \beta} \frac{\sin \pi \alpha}{\pi \alpha} t^{\alpha - \beta
+1}.
\end{equation}
For $\alpha < \beta -1$ and for $t$ large the behavior is different, Eq.(\ref{eq:AB2}) and the MSD stagnates:
\begin{equation}
  \langle \langle n_{1}( \Delta t ) \rangle \rangle \simeq C \left( \frac{\tau_0}{t_0} \right)^\alpha 
\end{equation}
with
\begin{equation}
 C= \frac{\sin \pi \alpha}{\pi \alpha} (\beta -1) \mathrm{B}(\alpha +1, \beta - \alpha -1).
\end{equation}
The stagnant number of steps is large only if $\tau_0 \gg t_0$. Only in this case any universal behavior of the PDF can be anticipated. \\

On the other hand, \textbf{for the incomplete resetting} we substitute Eq.~(\ref{nal1}) and Eq.~(\ref{p2b1}) into Eq.~(\ref{n2}) and introduce new variables $z=\tau_0/t$ and $\zeta= 1-t_r/t$:
\begin{equation}\label{n2w}
\langle \langle n_2(t) \rangle \rangle = \frac{\sin \pi \alpha}{\pi \alpha} \left(\frac{t}{t_0} \right)^\alpha \left(1 -  t^ {1 -\beta}\frac{\beta-1}{\tau_0^{1-\beta}}I_0 \right). 
\end{equation}
The integral 
\begin{equation}
I_0=\int_0^1 (1-\zeta)^\alpha (z + \zeta)^{-\beta} d\zeta
\end{equation}
diverges at lower limit for $z \to 0$. Close to this limit the first multiplier in the integrand can be set to unity and therefore 
\begin{equation}
I_0\simeq \int_0^1 (z + \zeta)^{-\beta} d\zeta \simeq \frac{z^{1-\beta}}{\beta-1} 
= \frac{\tau_0^{1 - \beta}}{t^{1-\beta} (\beta -1)}.
\end{equation}
In contrast to the case $\beta < 1$,the second term in the brackets in Eq.~(\ref{n2w}) converges to unity for $t \to \infty$, and the main asymptotics of the expression comes from subleading terms. The reason is that for $\beta > 1$ the PDF $p(t_r|t)$ is very strongly peaked at $t_r \approx t$, and the difference between $t_r$ and $t$ is typically small. 

The way to circumvent the calculation of the subleading terms is as follows. Introducing $\Delta t = t-t_r$ we now may expand the expression
Eq.~(\ref{eq:MMDeltat}) with substitution from Eq.~(\ref{nal1}) in $\Delta t$ and write
\begin{equation}
\langle n_2( \Delta t ) \rangle\simeq  \frac{\sin \pi \alpha}{\pi} t_0^{-\alpha} t^{\alpha-1} \Delta t.
\end{equation}
Performing the average this over the distribution of $\Delta t$, Eq.~(\ref{p1b1}) and introducing the variable $z=\tau_0/t$, we get
\begin{equation}
\langle \langle n_2(t) \rangle \rangle =
 \frac{\sin \pi \alpha}{\pi} \frac{t^{\alpha -1}}{t_0^\alpha} \frac{\beta-1}{\tau_0^{1-\beta}} t^{2 - \beta} I_1(1,\beta;z).
 \label{eq:steps2}
\end{equation}
where the integral $I_1$ is defined in terms of Eq.~(\ref{I1}). According to Eq.(\ref{eq:AB1}) we thus get for $\beta < 2$
\begin{equation}
  \langle \langle n_2(t) \rangle \rangle = \frac{\sin \pi \alpha}{\pi} t_0^{-\alpha} \tau_0^{\beta-1} \frac{\beta-1}{2-\beta} t^{1+\alpha - \beta}.
\end{equation}
Depending on the relation between $\alpha$ and $\beta$ this may be a decaying or a growing function of $t$. Thus, for $\alpha > \beta -1$, $\langle \langle n(t) \rangle \rangle$
grows at longer times
monotonically, and the typical number of steps will be large. In the opposite case the number of steps would decay at longer times, and can be large only in the intermediate time
domain
\begin{equation}
t \ll (t_0^{\alpha} \tau_0^{\beta-1})^\frac{1}{\beta - 1-\alpha} = \tau_0 \left(\frac{\tau_0}{t_0} \right)^{\frac{\alpha}{\beta - 1 -\alpha}}.
\end{equation}
Noting that our asymptotic discussion is only valid for $t_0,\tau_0 \ll t$, the necessary condition of the existence of large $\langle \langle n(t) \rangle \rangle$ is
\begin{equation}
 t_0,\tau_0 \ll \tau_0 \left(\frac{\tau_0}{t_0} \right)^\frac{\alpha}{\beta - 1-\alpha} 
\end{equation}
which would hold for $t_0 \ll \tau_0$. \\

For $\beta > 2$ we have 
\begin{equation}
\frac{\sin \pi \alpha}{\pi} \frac{t^{\alpha -1}}{t_0^\alpha} \frac{\beta-1}{\tau_0^{1-\beta}} t^{2 - \beta} z^{2-\beta} \mathrm{B}(2,\beta - 2) = C_1 \frac{t^{\alpha
-1}}{t_0^\alpha \tau_0},
\end{equation}
which is a decaying function of $t$. To get the intermediate domain in which $\langle \langle n(t) \rangle \rangle \gg 1$ together with
$t_0, \tau_0 \ll t$ one again needs to chose $\tau_0 \gg t_0$. \\

\subsubsection{Normal diffusion with $\alpha \geq 1$ }

For the case $\alpha > 1$ we have for \textbf{complete resetting}
\begin{equation}
 \langle \langle n_{1}(t) \rangle \rangle = \frac{\alpha -1}{t_0}\frac{\beta - 1}{\tau_0^{1-\beta}} \int_0^t \frac{\Delta t}{(\tau_0 + \Delta t)^\beta} d\Delta t. 
\end{equation}
Changing the variable of integration to $\xi = \Delta t/t$ and taking $z= \tau_0/t$ leads to
\begin{equation}
 \langle \langle n_{1}(t) \rangle \rangle = \frac{\alpha -1}{t_0}\frac{\beta - 1}{\tau_0^{1-\beta}} t^{2 - \beta} I_1(1, \beta; z)\,,
\end{equation}
where the integral $I_1$ is defined in terms of Eq.~(\ref{I1}). The result depends on whether $1 < \beta < 2$ or $\beta > 2$. For $\beta < 2$ Eq.(\ref{eq:AB1}) applies with
\begin{equation}
 \langle \langle n_{1}(t) \rangle \rangle = \frac{\alpha -1}{t_0}\frac{\beta - 1}{(3-\beta)\tau_0^{1-\beta}} t^{2 - \beta} .
\end{equation}
For $\beta > 2$ we have for $z \to 0$ 
\begin{equation}
\langle \langle n_{1}(t) \rangle \rangle = \mathrm{B}(2,\beta - 2) (\alpha -1)(\beta - 1) \frac{\tau_0}{t_0},
\end{equation}
i.e. $\langle \langle n_{1}(t) \rangle \rangle $ tends to a constant which is large provided $\tau_0 \gg t_0$.\\

For the \textbf{incomplete resetting} we obtain the same result, $\langle \langle n_{2}(t) \rangle \rangle=\langle \langle n_{1}(t) \rangle \rangle$.

\begin{figure}[tbp]
\begin{center}
\scalebox{0.4}{\includegraphics{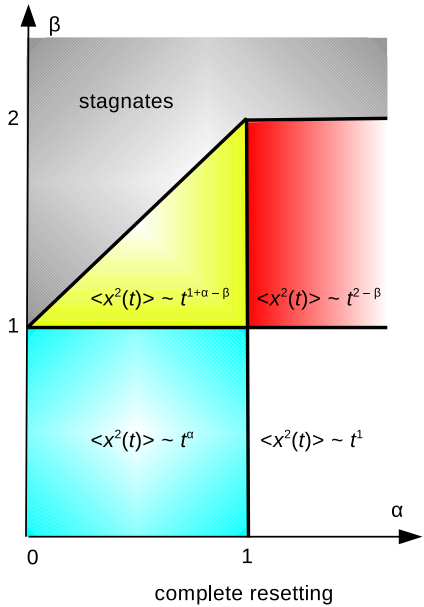} } \\
\scalebox{0.4}{\includegraphics{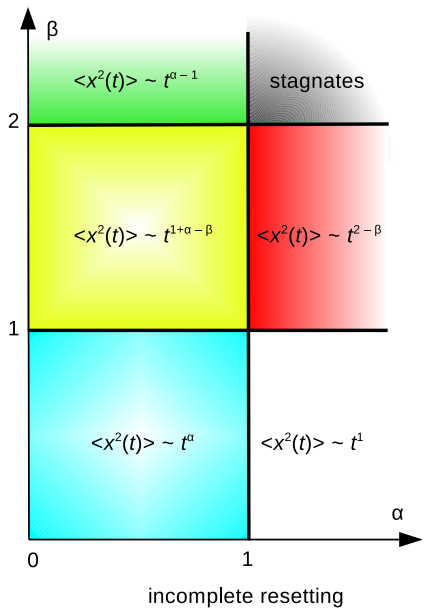} }
\caption{(Color online) The time dependence of the MSD in different domains of parameters $\alpha$ and $\beta$ for 
complete and for incomplete resetting. Note that these dependencies are the same as for the mean-filed model, the scaled Brownian motion (SBM) \cite{AnnaNonrenewal,AnnaRenewal}.
The case of complete resetting corresponds to the renewal \cite{AnnaRenewal}, and the case of incomplete resetting to the non-renewal \cite{AnnaNonrenewal} cases for the SBM.
\label{fig:MSD}}
\end{center}
\end{figure}

\subsection{Asymptotic of the mean number of steps}

The Table \ref{tab:3} represents the time domains in which the mean number of steps is much larger than unity. Here the notation is as follows:
If $\langle \langle n(t) \rangle \rangle \to \infty$ in the limit $t \to \infty$, the behavior is called asymptotic. In other cases $\langle \langle n(t) \rangle \rangle \gg 1$
only when $\tau_0 \gg t_0$. This may take place at any value of $t$ provided it is large enough, $t \gg t_0, \tau_0$ or only in some
domain of $t$ bounded from above. In the first case we will say that the behavior is independent of $t$, and in the second case that the behavior is transient. 
These results will be of use in Section IV.

\begin{table}[h!]
\caption{Conditions for $\langle \langle n(t) \rangle \rangle \gg 1$. \label{tab:3}}
\begin{center}
 \begin{tabular}{|c|c|c|c |} \hline
$\beta$ & $\alpha$ & complete & incomplete \\
 & & resetting & resetting \\
\hline \hline
$0< \beta < 1$ & all $\alpha$ & $t \gg \tau_0, t_0$ & $t \gg \tau_0, t_0$ \\
& &  (asymptotic) & (asymptotic) \\
\hline
$1 < \beta < 2$ & $\alpha > \beta - 1$ & $t \gg \tau_0, t_0$  & $t \gg \tau_0, t_0$  \\
& &  (asymptotic) & (asymptotic) \\
\cline{2-4}
 & $ \alpha < \beta - 1$ & $\tau_0 \gg t_0$ &  $\tau_0 \gg t_0$  \\
 &  &  (all $t$) & $\tau_0 \ll t \ll \tau_0 \left(\frac{\tau_0}{t_0} \right)^{\frac{\alpha}{\beta - 1 - \alpha}} $\\
 & &  & (transient) \\
 \hline
 $ 2 < \beta$ & $\alpha < 1$ & $\tau_0 \gg t_0$ &  $\tau_0 \gg t_0$  \\
  & &  (all $t$) & $\tau_0 \ll t \ll \tau_0 \left(\frac{\tau_0}{t_0} \right)^{\frac{\alpha}{1 - \alpha}}$ \\
 & &  & (transient) \\
 \cline{2-4}
 & $\alpha > 1$ & $\tau_0 \gg t_0$ & $\tau_0 \gg t_0$ \\ 
 & &  (all $t$) & (all $t$) \\
 \hline
\end{tabular}
\end{center}
\end{table}

\subsection{Mean squared displacement: numerical results}

After $ \langle \langle n_{1,2}(t) \rangle \rangle$ are found, the behavior of the MSD $\langle x^2 (t) \rangle$ follows from Eq.(\ref{eq:MSDCTRW}).  The overview of all possible regimes
of the MSD is provided in Fig. \ref{fig:MSD}.

\begin{figure}
\begin{center}
\scalebox{0.3}{\includegraphics{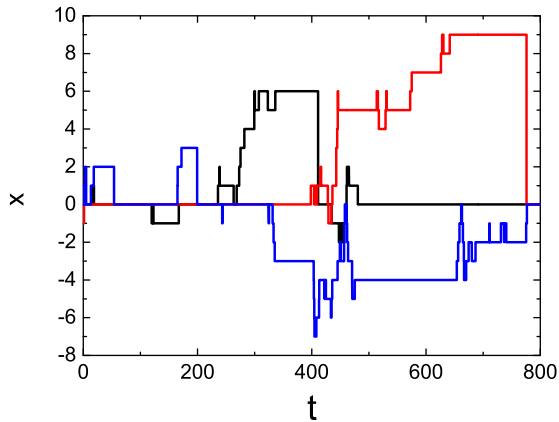} }
\caption{Typical trajectories for CTRW with power-law waiting time density for jumps and the power-law distribution of 
waiting times for resetting (incomplete resetting case). The parameters are $\beta=0.5$, $\alpha=0.5$, $\tau_0=1$. \label{Gxyzpowres} } 
\end{center}
\end{figure}

The analytical results for the MSD are confirmed in terms of the numerical simulation. For each realization of the process we generate random numbers $s_i$ distributed according to Eq.~(\ref{eq:CTRWpar}) for the CTRW
waiting times and random numbers $r_i$ distributed according to Eq.~(\ref{eq:respar}) for the resetting waiting times. We take $t_0=1$, corresponding to $K_\alpha = 0.318$  (according to Eq.~\ref{eq:DiffKoeff}). The values of $\tau_0$ differ in different simulations and are 
given explicitly in the captions or in the legends. The times of steps are then obtained as $t_1 = s_1$, $t_n = t_{n-1} + s_n$, and the procedure is stopped when $t_n$ exceeds the maximal simulation time $T$. 
The resetting times $r_n$ are generated in a similar manner. In the simulation of the CTRW the time, starting from $t=0$, is increased by an amount of $\Delta t$, and it is checked, 
whether a jump, or the resetting event falls in the corresponding time interval. In the first case the walker performs the jump with the
length $\Delta x = 1$  either to the right or to the left with equal probability. In the second case the coordinate of the walker is set to zero.
Fig.~\ref{Gxyzpowres} displays three trajectories for the CTRW with power-law waiting time density and power-law resetting in the case of incomplete resetting.
The simulations reported in other figures are performed with $10^5$ walkers. 

\begin{figure}
\begin{center}
\scalebox{0.3}{\includegraphics{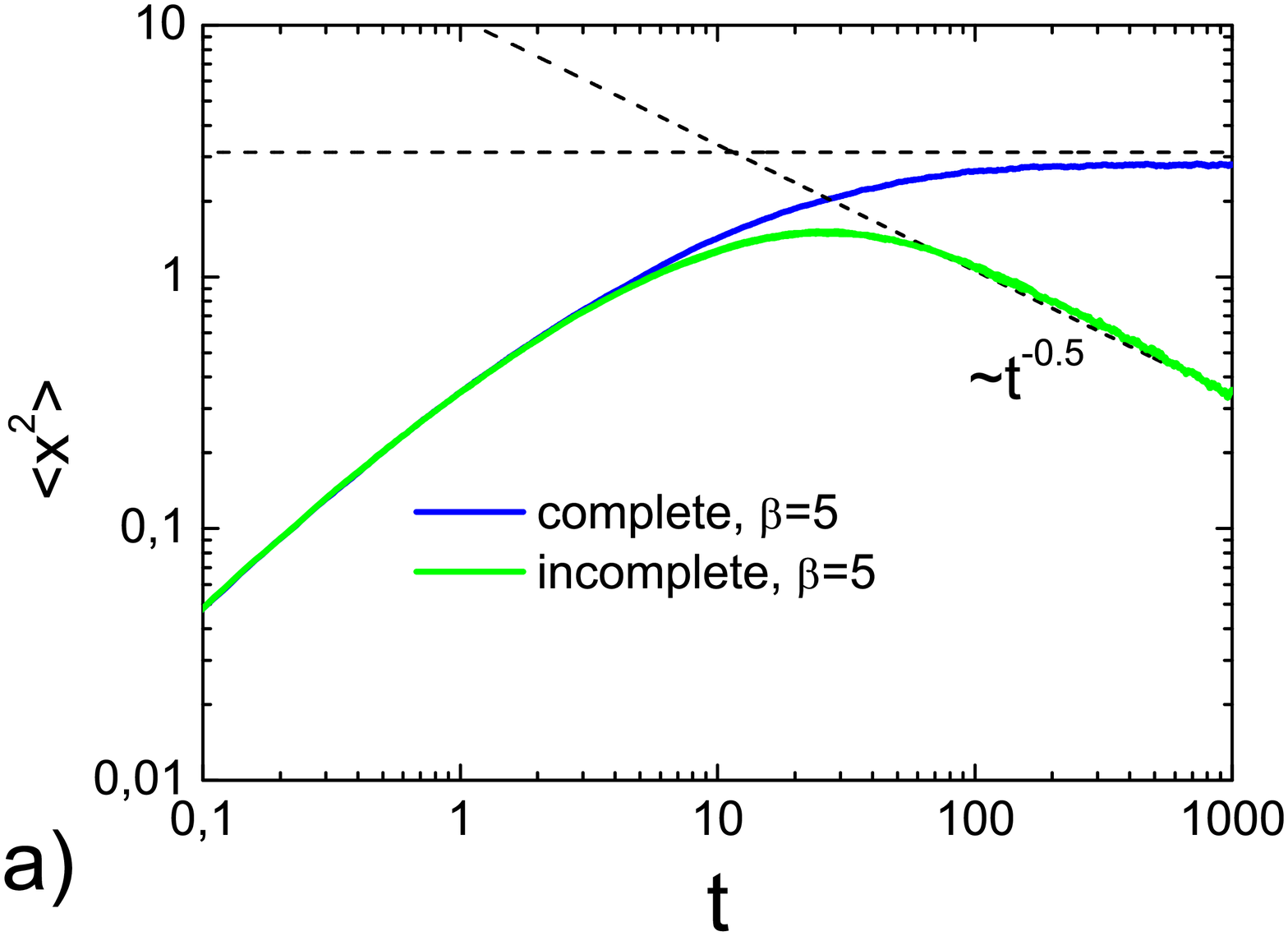}} \\
\scalebox{0.3}{\includegraphics{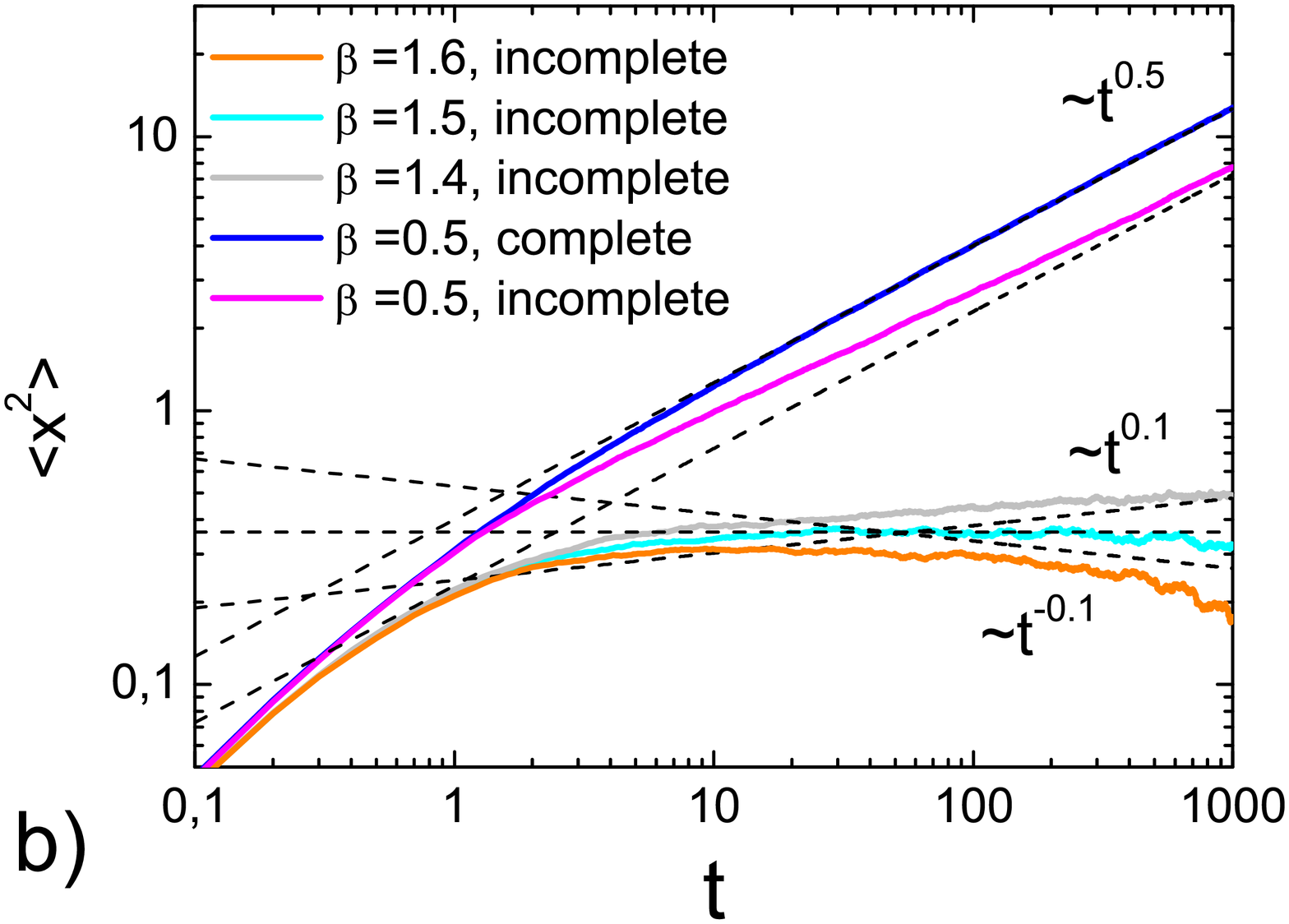}}
\caption{Mean squared displacement $\left\langle
x^2(t)\right\rangle$ for CTRW with the power law waiting time density with $\alpha=0.5$ and power-law resetting with $\tau_0=100$ and a) $\beta=5$, b) $\beta=0.5, 1.4,1.5$, and $1.6$. 
Dashed lines show the corresponding theoretical asymptotics. \label{Gpowpow}}
\end{center}
\end{figure}

In Fig.~\ref{Gpowpow} we show the simulation results for the MSD for both complete and incomplete resetting in a broad domain of parameters.
These simulations confirm that the corresponding asymptotics are the same as in the scaled Brownian motion, see Ref. \cite{AnnaRenewal} for
the renewal case, corresponding to complete resetting, and Ref. \cite{AnnaNonrenewal} for the non-renewal case (incomplete resetting). 

\section{Probability density functions}
\label{sec:PDFs}

\subsection{Asymptotic forms of the CTRW Green's functions}


The standard variant of the CTRW is the Scher-Montroll ``wait first'' scheme, starting from the waiting time. 
The PDF in the coordinate-time representation is 
\begin{equation}
 P_w(x,t) = \sum_{n=0}^\infty P_n(x)\chi_n(t)
\end{equation}
where $P_n(x)$ is the position after $n$ steps in a simple random walk, and $\chi_n(t)$ is the probability that exactly $n$ steps are taken up to the time $t$. 
The functions $\chi_n$ take a very simple form in the Laplace domain,
\begin{equation}
 \chi_n(s) = \psi^n(s) \frac{1-\psi(s)}{s},
\end{equation}
and the functions $P_n(x)$ in Fourier domain read $P_n(k) = \lambda^n(k)$, where $\lambda(k)$ is the characteristic function of the displacement distribution in a single step.
The PDF of the walker's position in the Fourier-Laplace representation 
for this scheme is given by 
\begin{equation}
 p_w(k,s) = \frac{1}{1 - \lambda(k) \psi(s)} \frac{1-\psi(s)}{s}.
\end{equation}
Another scheme, the ``jump first'' one, differs only in the fact that the walk starts not from a waiting time but from a jump at $t=0$, so that 
\begin{equation}
 P_j(x,t) = \sum_{n=0}^\infty P_{n+1}(x)\chi_n(t),
\end{equation}
and
\begin{equation}
 p_j(k,s) = \frac{\lambda(k)}{1 - \lambda(k) \psi(s)} \frac{1-\psi(s)}{s}.
\end{equation}
Assuming the steps to be symmetric and to have the finite second moment ($\lambda(k) \simeq 1 - a^2 k^2/2$) and the waiting times to follow a power law,
$\psi(s) \simeq 1 - \Gamma(1-\alpha) t_0^\alpha s^\alpha$ we get in both cases in the lowest order in $k$ and $s$ (i.e. in the continuous limit of long times and large scales) 
the same asymptotic expression 
\begin{equation}
 p(k,s) = \frac{s^{\alpha - 1}}{k^2 [a^2 / 2 \Gamma(1-\alpha)t_0^\alpha]+ s^\alpha}.
 \label{eq:GFasy}
\end{equation}
The combination $\tilde{K}_\alpha = a^2 / 2 \Gamma(1-\alpha)t_0^\alpha$ of the specific parameters of the walk is related to the coefficient of the anomalous diffusion $K_\alpha$ defined in
Eq.(\ref{eq:DiffKoeff}), $\tilde{K}_\alpha = \Gamma(\alpha) K_\alpha$. 
The limiting form of the Green's function of CTRW is given by the inverse Fourier-Laplace transform of Eq.(\ref{eq:GFasy}),
and reads 
\begin{equation}
G_{1,2}(x,t)=P_{w,j}(x,t) = \frac{1}{2 \sqrt{\tilde{K}_\alpha t^\alpha}} M_{\alpha / 2} \left(\frac{ |x|}{ \sqrt{\tilde{K}_\alpha t^\alpha} }\right). 
\label{eq:Green}
\end{equation}
with
\begin{equation}
 M_{\alpha / 2}(y) = \sum_{n=0}^\infty \frac{(-y)^n}{n! \Gamma[-\alpha n/2 + (1-\alpha/2)]}
\end{equation}
being the Mainardi function, see Appendix B.  
For $|x|$ large compared to $\sqrt{\tilde{K}_\alpha} t^{\alpha / 2}$ the function $G(x,t)$ shows a squeezed exponential tail, Eq.(\ref{eq:Masympt}).
For $|x|$ small compared to $\sqrt{\tilde{K}_\alpha} t^{\alpha / 2}$ the function $G(x,t)$ shows the cusp at zero
which disappears only in the Gaussian case $\alpha = 1$. The asymptotic form, Eq.(\ref{eq:Green}) applies when the number of steps performed during the time $t$ is large. 

Thus, for $|x| \ll K t^{\alpha/2}$ the Green's function tends to $G(0, t) = C_1 \Delta t^{-\alpha/2}$, while for $|x| \gg  K t^{\alpha/2}$ 
the leading asymptotics of the Green's function is 
\begin{equation}
 G(x,t) \simeq C_2 t^{-\frac{\alpha}{2(2-\alpha)}} |x|^\frac{2 \alpha - 2}{2 (2-\alpha)} \exp \left(C_3 \frac{|x|^{\frac{2}{2-\alpha}}}{t^{\frac{\alpha}{2-\alpha} }}\right).
\end{equation}
We will never need the exact form of the Green's function but only its similarity form, Eq.(\ref{eq:GFasy}), combined with the fact that the 
Mainardi function is rapidly decaying at infinity. 

According to our discussion accompanying Fig.1, the PDFs of the CTRW under resetting $P(x,t)$ is given by mixtures of the PDFs (Green's functions) of the CTRW $G_{1,2}(x,\Delta t)$
of the
wait first or jump first CTRW in cases (1) and (2), respectively. These Green's functions are weighted with the 
probability density $p(\Delta t |t)$ of the observed duration of the corresponding walk $\Delta t$, conditioned on the observation time $t$ in the case (1) or with the probability
density of the time $\Delta t'$ elapsed between the first step of the walk after the resetting and the end of the observation, $p_2(\Delta t' | t)$ (\textit{vide infra}).
Thus,
\begin{equation}
 P_{1}(x,t) = \int_0^t G_{1}(x,\Delta t) p_1(\Delta t | t) d \Delta t
 \label{eq:General1}
\end{equation}
and 
\begin{equation}
 P_{2}(x,t) = \int_0^t G_{2}(x,\Delta t') p_{2}(\Delta t' | t) d  \Delta t'.
 \label{eq:General_WF}
\end{equation}
Let us assume that the PDF $p(\Delta t | t) = p_{1,2}(\Delta t | t)$ of the CTRW duration $\Delta t$ (or $\Delta t'$) follows a power law
in some domain of $\Delta t$, i.e. possesses an intermediate asymptotics
\begin{equation}
 p(\Delta t | t) = A(t) \Delta t^{-\gamma}
 \label{eq:funct}
\end{equation}
with $\gamma > 0$ in the domain $t_{\min} \ll \Delta t \ll t$. Then the corresponding intermediate asymptotics of $P(x, t)$
will be
\begin{eqnarray}
P(x, t) &=& A(t) \int_0^t d \Delta t \;  \Delta t^{-\gamma} \frac{1}{2 \sqrt{\tilde{K}_\alpha} \Delta t^{\alpha / 2}} \times \nonumber  \\
&& M_{\alpha / 2} \left(\frac{|x|}{\sqrt{\tilde{K}_\alpha} \Delta t^{\alpha / 2} } \right) .
\end{eqnarray}
Introducing the scaling variable
\begin{equation}
 \xi = \frac{|x|}{\sqrt{\tilde{K}_\alpha} \Delta t^{\alpha / 2}  }
\end{equation}
we rewrite the last expression as
\begin{eqnarray}
 P(x, t) &=& A(t) \tilde{K}_\alpha^{\frac{\gamma -1}{\alpha}} |x|^{-1 - \frac{2(\gamma -1)}{\alpha}} \times \nonumber  \\
&&  \int_{\frac{|x|}{\sqrt{\tilde{K}_\alpha} t^{\alpha/2}}}^\infty \xi^{\frac{2(\gamma - 1)}{\alpha}} M_{\alpha/2}(\xi) d\xi.
 \label{eq:General2}
\end{eqnarray}
The existence of the intermediate power-law asymptotics in $x$ (i.e. of the universal behavior for $|x| \ll \sqrt{\tilde{K}_\alpha} t^{\alpha / 2}$) corresponds to situations
when the integral stays convergent when its lower limit tends to zero, i.e. for $2(\gamma - 1)/\alpha > -1$, or, in other words, for
\begin{equation}
\gamma > 1 - \frac{\alpha}{2}.
 \label{eq:mainneq}
\end{equation}
In the opposite case the integral for small $|x|$ is dominated by its behavior on the lower limit of integration, where the Mainardi function tends to a 
constant, so that $P(x,t) \propto const \cdot A(t) t^{1 -\gamma - \alpha/2} x^0$, i.e. develops a flat top. 
Therefore the intermediate power-law asymptotics of the PDF exists for $\gamma > 1 - \alpha / 2$ and is given by
\begin{equation}
 P(x,t) \propto A(t) |x|^{-1 - \frac{2(\gamma -1)}{\alpha}}. 
 \label{eq:PrAs}
\end{equation}
The far asymptotics of large $|x|$ follows (up to power-law prefactors) the squeezed exponential wing of the Mainardi function.

\subsection{Equations for the distributions of observed walk duration}

The PDF of $\Delta t$ in the case of the complete resetting is given by Eq.(\ref{p1initial}) with its two special cases, Eqs.(\ref{p1b0}) and (\ref{p1b1}). 
The final results follow from the observations that for $\Delta t \ll t$ in the case (1) for $\beta < 1$
\begin{equation}
 p_1(\Delta t | t) \simeq \frac{\sin \pi \beta}{\pi} t^{\beta -1} \Delta t^{-\beta}
\end{equation}
so that $A(t) \sim t^{\beta -1}$ and  $\gamma = \beta$, and 
\begin{equation}
 p_1(\Delta t | t)  \simeq \frac{\beta - 1}{\tau_0^{1-\beta}} \Delta t^{-\beta}
\end{equation}
so that $A(t) = const \cdot \tau_0^{\beta -1}$ and  $\gamma = \beta$ for $\beta > 1$. \\

The distribution of the duration $\Delta t'$ of the jump-first CTRW in the case (2) of incomplete resetting was not considered yet. This 
CTRW starts after the forward recurrence time $t_f= t_1 - t_r$ after the resetting event, so that its duration is $\Delta t' = t - t_1$.  

Given $t_r$ (which is the aging time of the aged CTRW), the distribution of the forward recurrence time $t_f$ is given by \cite{sokbook}
\begin{equation}
 \psi_1(t_f|t_r) = \frac{\sin \pi \alpha}{\pi} \left(\frac{t_r}{t_f} \right)^\alpha \frac{1}{t_r+t_f}.
 \label{eq:psi1}
\end{equation}
The duration $\Delta t'$ of the following ``jump first'' CTRW is $\Delta t'= t -( t_r + t_f)$ if the sum $t_r + t_f$ does not exceed $t$, and is zero otherwise. 
Let us first fix $t_r$ and calculate the conditional PDF $p(\Delta t' | t_r, t)$:
\begin{eqnarray}
 p(\Delta t' | t_r,t) &=& \int_0^{t-t_r} dt_f \delta[\Delta t' -(t - t_r-t_f)] \psi_1(t_f|t_r) \nonumber \\
 && + \delta(\Delta t') \int_{t-t_r}^\infty \psi_1(t_f|t_r) dt_f \nonumber \\
&=& \frac{\sin \pi \alpha}{\pi} \left(\frac{t_r}{t - t_r - \Delta t'} \right)^\alpha \frac{1}{t - \Delta t'} \nonumber \\
&& + \delta(\Delta t') \int_{t-t_r}^\infty \psi_1(t_f|t_r) dt_f,
\label{eq:cond1}
\end{eqnarray}
where $\delta(x)$ is the Dirac delta function. The weight of this $\delta$-term, the integral $I(t-t_r) = \int_{t-t_r}^\infty \psi_1(t_f|t_r) dt_f$, is the probability that 
no steps of CTRW were done after resetting. Now we average the expression Eq.(\ref{eq:cond1}) over $t_r$ which has to lay between 0 and $t-\Delta t'$ if $\Delta t'$ is nonzero:
\begin{eqnarray}
&& p_2(\Delta t' | t) = \int_0^{t-\Delta t'} p(\Delta t' | t_r,t) p_r(t_r | t) dt_r \nonumber \\
&& \qquad =\int_0^{t-\Delta t'} \frac{\sin \pi \alpha}{\pi} \left(\frac{t_r}{t - t_r - \Delta t'} \right)^\alpha \frac{1}{t - \Delta t'} p_r(t_r| t) dt_r \nonumber \\
 && \qquad + \delta(\Delta t') \int_0^t  \left[\int_{t-t_r}^\infty \psi_1(t_f|t_r) dt_f \right] p_r(t_r|t) dt_r.
 \label{eq:p2}
\end{eqnarray}
The term with zero steps contributes to the overall normalization and corresponds to a delta-peak at the origin in the total PDF. This term does not influence 
the wings of the PDF. 
We will denote the weight of the $\delta$-function in the last line by $R$.

The explicit form of $R$ for \boldmath $\beta < 1$ \unboldmath is
\begin{eqnarray}
 R &=& \frac{\sin \pi \alpha}{\pi }\frac{\sin \pi \beta}{\pi } \times \label{eq:rest} \\
 && \int_0^t t_r^{\beta -1}(t-t_r)^{-\beta}\int_{t-t_r}^\infty t_r^\alpha t_f^{-\alpha} (t_r+t_f)^{-1} dt_r dt_f. \nonumber
\end{eqnarray}
We note that the conditional PDF $\psi_1(t_f|t_r)$, Eq.(\ref{eq:psi1}), is normalized for any $t_r$, and therefore $\int_0^t p_r(t_r|t) \left[ \int_0^\infty \psi_1(t_f |t_r) dt_f
\right] dt_r = 1$.
Note that the integrand of the second integral in Eq.(\ref{eq:rest}) is non-negative, so that 
\begin{equation}
 \int_{t-t_r}^\infty t_r^\alpha t_f^{-\alpha} (t_r+t_f)^{-1} dt_f \leq \int_{0}^\infty t_r^\alpha t_f^{-\alpha} (t_r+t_f)^{-1} dt_f 
\end{equation}
and therefore $R \leq 1$, so that the whole double integral has to be convergent (except for the limiting cases 
$\alpha =1$ or $\beta = 1$ when the trigonometric prefactors vanish). On the other hand, introducing the new variables $\xi = t_r/t$ and $\eta = t_f/t$ we see that
\begin{eqnarray}
R &=& t^0 \frac{\sin \pi \alpha}{\pi }\frac{\sin \pi \beta}{\pi } \times\\
&& \int_0^1 \xi^{\beta -1} (1-\xi)^{-\beta} \left[ \int_{1-\xi}^\infty \xi^\alpha \eta^{-\alpha} (\xi + \eta)^{-1} d\eta \right] d \xi.  \nonumber 
\end{eqnarray}
The integral in this expression converges, as we have seen above, is positive, and depends only on parameters $\alpha$ and $\beta$, but not on $t$. 
Therefore the weight of the $\delta$-peak tends to a constant in the course of time. 

For \boldmath $\beta > 1$ \unboldmath the qualitative result is the same, but the discussion is slightly different. Now 
\begin{equation}
 p_r(t_r | t) \simeq \frac{1}{\beta - 1} \frac{\tau_0^{\beta -1}}{(\tau_0 + t - t_r)^\beta}, 
\end{equation}
so that 
\begin{eqnarray}
  R &=&  \frac{1}{\beta - 1} \frac{\sin \pi \alpha}{\pi } \tau_0^{\beta -1} \times \\
  && \int_0^t (\tau_0 + t-t_r)^{-\beta}\int_{t-t_r}^\infty t_r^\alpha t_f^{-\alpha} (t_r+t_f)^{-1} dt_r dt_f. \nonumber 
\end{eqnarray}
Denoting $\zeta = \tau_0/t$ we write
\begin{eqnarray}
R &=&  \frac{1}{\beta - 1} \frac{\sin \pi \alpha}{\pi} \zeta^{\beta -1} \times  \\
&& \int_0^1 (1+\zeta - \xi)^{-\beta} \int_{1-\xi}^\infty \xi^\alpha \eta^{-\alpha}(\xi + \eta)^{-1} d\eta d \xi . \nonumber 
\end{eqnarray}
Now we note that this expression is bounded from above (since $R \leq 1$) and would tend to zero only if the double integral 
in the last expression converges or diverges slower that $\zeta^{1-\beta}$ for $\zeta \to 0$. 
Now we introduce the new variable $z=  \eta/\xi$ in the inner integral, and write
\begin{eqnarray}
 R &=&  \frac{1}{\beta - 1} \frac{\sin \pi \alpha}{\pi} \zeta^{\beta -1} \times \\
 && \int_0^1 (1+\zeta - \xi)^{-\beta} \left[\int_{\xi^{-1}-1}^\infty z^{-\alpha} (1 + z)^{-1} d z\right] d \xi. \nonumber 
\end{eqnarray}
The $\zeta$-dependence of the whole integral is dominated by the behavior of the integrand for $\xi \to 1$ when the internal integral tends to a constant
\begin{equation}
 \int_0^\infty z^{-\alpha} (1 + z)^{-1} d z = \frac{\pi}{\sin [(1-\alpha) \pi ] },
\end{equation}
and therefore 
\begin{eqnarray}
&& \int_0^1 (1+\zeta - \xi)^{-\beta} \int_{1-\xi}^\infty \xi^\alpha \eta^{-\alpha}(\xi + \eta)^{-1} d\eta d \xi  \nonumber \\  
&& \qquad \simeq \frac{\pi}{(\beta - 1)\sin [(1-\alpha) \pi ] } \zeta^{1-\beta},
\end{eqnarray}
so that for $t \to \infty$
\begin{eqnarray}
 R &\to& \frac{1}{\beta - 1} \frac{\sin \pi \alpha}{\pi} \frac{\pi}{(\beta - 1)\sin [(1-\alpha) \pi ] }  \nonumber \\
 &=& \frac{1}{(\beta -1)^2} \frac{\sin \pi \alpha}{\sin \pi(1-\alpha)},
\end{eqnarray}
i.e. again tends to the constant. The $\delta$-peak only disappears for $\beta \to \infty$ and for exponential resetting. 

The main integral (the second line in Eq.(\ref{eq:p2})) is awkward, but we can still distill the general time dependence (up to prefactors).  
To do so we note that the intermediate asymptotics appears when for $t_0 \ll \Delta t' \ll t$ the function $p_2(\Delta t' | t)$ possesses a power-law asymptotics 
$p_2 \sim A(t)\Delta t'^{-\gamma}$.

\paragraph{\boldmath $\beta < 1$ \unboldmath. } For this case we have 
\begin{eqnarray}
 && p(\Delta t' | t) = \int_0^{t-\Delta t'} p(\Delta t' | t_r,t) p_2(t_r | t) dt_r \nonumber \\
&& = \int_0^{t-\Delta t'} \frac{\sin \pi \alpha}{\pi} \left(\frac{t_r}{t - t_r - \Delta t'} \right)^\alpha \times \nonumber \\ 
&& \frac{1}{t - \Delta t'} \frac{\sin \pi \beta}{\pi} t_r^{\beta -1}(t-t_r)^{-\beta} dt_r \nonumber \\
&& + R \delta(\Delta t'). \label{eq:NRGen}
\end{eqnarray}
Thus:
\begin{eqnarray}
p(\Delta t' | t) = && \frac{\sin \pi \alpha \sin \pi \beta}{\pi^2} \frac{1}{t - \Delta t'} \int_0^{t-\Delta t'} \times \nonumber \\
&& (t - \Delta t'- t_r)^{-\alpha} t_r^{\alpha+\beta -1}(t-t_r)^{-\beta} dt_r \nonumber \\
 && + R \delta(\Delta t'). 
 \label{eq:NR2}
\end{eqnarray}

The intermediate asymptotic power-law behavior in the wing of the PDF may appear if for $t$ long the PDF $p(\Delta t' | t)$ shows a power-law behavior for 
$t_0 \ll \Delta t' \ll t$, in which the $\delta$-peak does not play a role. To distill the power-law dependence on $\Delta t'$ we introduce in Eq.(\ref{eq:NR2}) new variables
$z=\Delta t'/t$ and $\xi=t_r/t$ and rewrite the integral
as 
\begin{eqnarray}
&& p(\Delta t' | t) = \frac{\sin \pi \alpha \sin \pi \beta}{\pi^2} t^{-1} \frac{1}{1-z}  \times \\
&&  \qquad \qquad \int_0^{1-z} (1  -z- \xi)^{-\alpha}   \xi^{\alpha+\beta -1} (1-\xi)^{-\beta} d\xi. \nonumber 
\end{eqnarray}
Now we investigate the behavior of the integral for $z \to 0$. This behavior depends on whether $\alpha + \beta < 1$ or $\alpha + \beta > 1$.
In the first case the integral converges and tends to a constant value. This corresponds to $\gamma = 0$. In the second case it shows a divergence at its upper limit.
Since this limit is approximately unity, we can set the second multiplier in the integrand to unity and simplify the expression:
\begin{equation}
 p(\Delta t' | t) \simeq C \times t^{-1}  \int_0^{1-z} (1  -z- \xi)^{-\alpha} (1-\xi)^{-\beta} d\xi.
\end{equation}

Now we introduce the new variable of integration $\zeta = 1 - z - \xi$ and write:
\begin{eqnarray}
&& I=\int_0^{1-z} (1  -z- \xi)^{-\alpha} (1-\xi)^{-\beta} d\xi \\
&& \;\;\;  = \int_0^{1-z} \zeta^{-\alpha}(z+\zeta)^{-\beta} d \zeta  \nonumber \\
&& \;\;\;  = \frac{1}{1-\alpha}(1-z)^{1-\alpha} z^{-\beta} \;_2F_1 \left(\beta, 1-\alpha, 2-\alpha;-\frac{1-z}{z} \right), \nonumber 
\end{eqnarray}
see Eq.(1.2.4.3) of Ref. \cite{BrPr}. Now we apply the Pfaff transformation 
\begin{eqnarray}
 && 2F_1 \left(\beta, 1-\alpha, 2-\alpha; x \right) \\
 && \qquad = (1-x)^{-1+\alpha}\;_2F_1 \left(1-\alpha,2-\alpha- \beta; 2-\alpha; \frac{x}{x-1} \right),  \nonumber 
\end{eqnarray}
so that (for $z \to 0$)
\begin{equation}
 I \to  \frac{1}{1-\alpha} z^{1-\alpha - \beta} \;_2F_1(1-\alpha, 2-\alpha - \beta; 2-\alpha;1) \sim z^{1-\alpha - \beta} .
\end{equation}
The value of the corresponding hypergeometric function is 
\begin{equation}
 \;_2F_1(1-\alpha, 2-\alpha - \beta; 2-\alpha;1) = \frac{\Gamma(2-\alpha)\Gamma(\alpha + \beta -1)}{\Gamma(1)\Gamma(\beta)}
\end{equation}
(note that $\alpha + \beta -1 >0$ is exactly the condition under which this asymptotic value is attained), so that
\begin{eqnarray}
&& p(\Delta t' | t) \simeq \frac{\sin \pi \alpha \sin \pi \beta}{\pi^2} \frac{\Gamma(2-\alpha)\Gamma(\alpha + \beta -1)}{\Gamma(1)\Gamma(\beta)}\times \nonumber \\
&& \qquad \qquad t^{-1} \left(\frac{\Delta t'}{t} \right)^{1 - \alpha - \beta},
\end{eqnarray}
which corresponds to our power law with $A(t) \propto t^{-\alpha - \beta}$ and $\gamma = \alpha + \beta -1$. 

\paragraph{\boldmath $ \beta > 1$ \unboldmath.} For the case $\beta > 1$ we have 
\begin{eqnarray}
p_2(\Delta t' |t) &\simeq& \int_0^{t-\Delta t'} \frac{\sin \pi \alpha}{\pi} \left(\frac{t_r}{t - t_r - \Delta t'} \right)^\alpha \times  \nonumber \\
&& \frac{1}{t - \Delta t'} \frac{1}{\beta - 1} \frac{\tau_0^{\beta -1}}{(\tau_0 + t - t_r)^\beta} dt_r.
\end{eqnarray}
Now we again introduce $z= \Delta t'/t$, $\zeta = \tau_0/t$ and $\xi = t_r/t$ and obtain
\begin{eqnarray}
p_2(\Delta t' |t) &\simeq& t^{-1} \frac{\sin \pi \alpha}{\pi (\beta - 1)} \frac{\zeta^{\beta -1}}{1-z} \times \nonumber \\
&& \int_0^{1-z} \frac{\xi^\alpha}{(z+1-\xi)^{\alpha}(\zeta + 1 - \xi)^\beta} d \xi.
\end{eqnarray}
We are interested in the asymptotic $z$-dependence of this expression for $z \to 0$. We note that at $z=0$ the integral stays convergent, 
however the interesting condition is $z \gg \zeta$. For both $z$ and $\zeta$ small the integral is dominated by the behavior of the integrand at the upper bound,
where, due to the restriction $z \gg \zeta$ we can neglect $\zeta$ in the second multiplier in the denominator, take $\xi^\alpha \approx 1$ in the numerator and change the
integration variable to $y=1-\xi$:
\begin{eqnarray}
I &=&  \int_0^{1-z} \frac{\xi^\alpha}{(z+1-\xi)^{\alpha}(\zeta + 1 - \xi)^\beta} d \xi \nonumber  \\
&\approx&  \int_z^1 (z+y)^{-\alpha}(\zeta + y)^{-\beta} d y.
\end{eqnarray}
Now, close to the lower bound the first term is simply $z^{-\alpha}$ and in the second one the small regularizing term $\zeta$ can be neglected, so that
\begin{equation}
 I \approx z^{-\alpha} \int_z^1 y^{-\beta} dy \simeq \frac{1}{\beta -1} z^{1-\alpha -\beta}. 
\end{equation}
Putting this in the expression for $p_2$ we get
\begin{eqnarray}
p_2(\Delta t' |t) &\simeq& t^{-1} \frac{\sin \pi \alpha}{\pi (\beta - 1)^2} \zeta^{\beta -1} z^{1-\alpha -\beta}  \nonumber \\
&=& \frac{\sin \pi \alpha}{\pi (\beta - 1)^2} \tau_0^{\beta - 1} t^{\alpha-1} \Delta t'^{1-\alpha -\beta},
\end{eqnarray}
and get our power law expression with $A(t) \propto \tau_0^{\beta - 1} t^{\alpha-1}$, and $\gamma = \alpha + \beta -1$ like in the previous case. 

The overall results for the intermediate asymptotics pf $P_{1,2}(\Delta t' | t)$ are summarized in Table \ref{tab:1}.
\\

\begin{table}[h!]
\caption{Intermediate asymptotics of random walk duration \label{tab:1}}
\begin{center}
 \begin{tabular}{|c|c|c|l|} \hline
kind of resetting & $A(t)$ & $\gamma$ & restrictions \\
\hline \hline
complete & $t^{\beta -1}$ & $\beta$ & $\beta < 1$ \\
\cline{2-4}
 &$\tau_0^{\beta -1}$ & $\beta$ & $\beta > 1$ \\
 \hline
 \hline
 & $t^{-1}$ & 0 & $\beta < 1$, $\alpha + \beta < 1$ \\
 \cline{2-4}
 incomplete & $t^{-\alpha - \beta}$ & $\alpha + \beta -1$ & $\beta < 1$, $\alpha + \beta > 1$ \\
  \cline{2-4}
 & $\tau_0^{\beta -1} t^{\alpha -1}$ & $\alpha + \beta - 1$ & $\beta > 1$ \\
 \hline
\end{tabular}
\end{center}
\end{table}

The inspection of this table allows us to tell under which conditions we can await the power-law intermediate behavior of the PDF, when we remember that this only appears for 
$\gamma > 1 - \alpha/2$, Eq.(\ref{eq:mainneq}). Thus, for $\beta < 1$ the intermediate asymptotics in the complete resetting is observed only for $\beta > 1-\alpha/2$, otherwise
the flat top of the PDF immediately merges with its squeezed exponential tail. For incomplete resetting with $\alpha + \beta < 1$ it does not exist at all (one has a delta-peak connected to
the wing), and for $\alpha + \beta >1$ the condition to observe the intermediate asymptotics is $\beta > 2 - \frac{2}{3} \alpha$ (under which condition the inequality $\alpha + \beta >1$
holds automatically for all $\alpha < 1$). 

\begin{figure}[h!]
\begin{center}
\scalebox{0.4}{\includegraphics{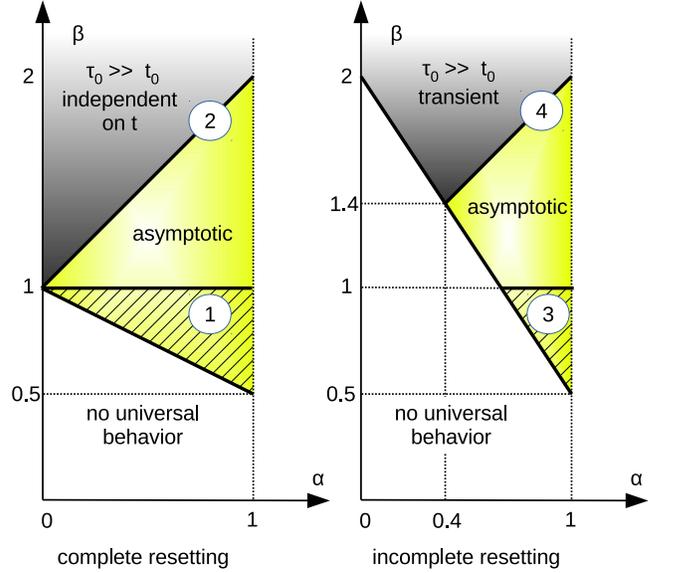} }
\caption{(Color online) Classification of intermediate asymptotic behaviors of $P(x,t)$ as a function of $x$ for CTRW under resetting for $0< \alpha < 1$. The lower, white, region
corresponds to the domain of parameters $\alpha, \beta$ where the intermediate power-law behavior does not occur. The intermediate triangular domain (yellow online) corresponds
to the values of parameters when the power-law behavior in $|x|$ is observed at long times. The gray domains correspond to the cases when this behavior is observed only 
for $\tau_0 \gg t_0$ either at all times (left) or only transiently (right). The hatched domains correspond to situations when the intermediate $|x|$ asymptotics 
is observed for $\beta < 1$. The types of behavior are: domain 1: $p(x,t) \sim t^{\beta -1} |x|^{-1 - \frac{2(\beta -1)}{\alpha}}$, domain 2: $p(x,t) \sim |x|^{-1 - \frac{2(\beta
-1)}{\alpha}}$,
domain 3: $p(x,t) \sim t^{-\alpha - \beta} |x|^{\frac{2(1-\beta)}{\alpha} - 3}$, and domain 4: $p(x,t) \sim \tau_0^{\beta -1} t^{\alpha - 1} |x|^{\frac{2(1-\beta)}{\alpha} - 3}$.
\label{fig:Domains}}
\end{center}
\end{figure}

\subsection{Final results}

For a complete resetting the final results are as follows: The intermediate asymptotics exists for $\beta > 1 - \frac{\alpha}{2}$, and reads
\begin{equation}
 P(x,t) \sim \left\{
 \begin{array}{lll}
  t^{\beta -1} |x|^{-1 - \frac{2(\beta -1)}{\alpha}} \qquad &\mbox{for } \beta < 1\\
  |x|^{-1 - \frac{2(\beta -1)}{\alpha}} \qquad &\mbox{for } \beta > 1.
 \end{array} 
 \right. 
 \label{eq:Fin_complete}
\end{equation}
This behavior is exactly the same as for SBM with the corresponding exponent of the anomalous diffusion $\alpha$, see Ref. \cite{AnnaRenewal}.
Note that for $\beta > 1$ and $\alpha < \beta -1$ the universal form of the PDF is only transient (i.e. visible only at intermediate times) and only exists for $\tau_0 \gg t_0$.

For incomplete resetting the intermediate asymptotics is visible only for $\beta > 2 - \frac{3}{2}\alpha$ and reads
\begin{equation}
 P_2(x,t) \sim \left\{
 \begin{array}{lll}
  t^{-\alpha - \beta} |x|^{\frac{2(1-\beta)}{\alpha} - 3} & \mbox{for  }& \beta < 1 \\
  \tau_0^{\beta -1} t^{\alpha - 1} |x|^{\frac{2(1-\beta)}{\alpha} - 3} & \mbox{for  }& \beta > 1 .
 \end{array}
 \right. 
\end{equation}
We have to stress that the universal form of the Green's functions based on taking only the lowest order contribution in $k$ is only applicable if the corresponding PDF is broad
enough, i.e.
typical value of $|x|$ is much larger than $a$. This implies that the number of steps of CTRW made during the time $t$ must be large. The typical number of steps is of the order 
of $\langle \langle n(t) \rangle \rangle$, whose behavior was already discussed in Sec. \ref{sec:MSD}.

Note that for incomplete resetting the corresponding behavior represents a decaying function of $|x|$, which is ``switched'' between the 
delta-peak at the origin and the squeezed exponential tail, starting late. This behavior differs from the one observed in SBM both
with respect to the existence of the $\delta$-peak and with respect to the presence of the $\alpha$-dependence in the corresponding power law. Both features
are connected with the fact that the first step of the CTRW after resetting follows very late after the resetting event. This is a true fluctuation
effect, which is not captured by the mean-field SBM-description. Note that since the prefactor of $|x|$ explicitly depends on time
the situation is always nonstationary. In this case again the universal asymptotic behavior in the case $\alpha < \beta -1$ is only observable for $t_0 \ll \tau_0$. 

\begin{figure}
\begin{center}
\scalebox{0.3}{\includegraphics{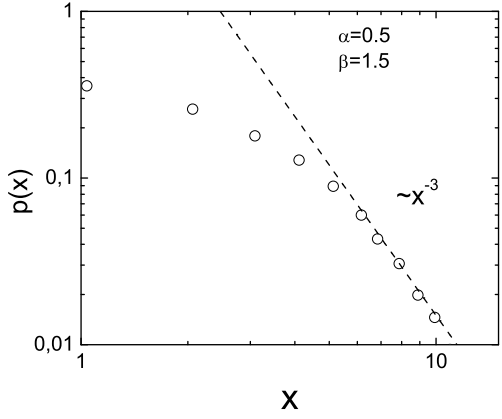}}
\caption{Probability density function for CTRW with power-law densities for waiting times and for resetting times under complete resetting.
The parameters are: $\alpha=0.5, \beta=1.5$, $t_0=1$ and $\tau_0=1000$, $t=1000$. The wing of the distribution scales according to $p(x)\sim x^{-1-2\beta/\alpha+2/\alpha}$, corresponding to the domain 2 in Fig. \ref{fig:Domains}.\label{Gpdfc}} 
\end{center}
\end{figure}

\begin{figure}
\begin{center}
\scalebox{0.3}{\includegraphics{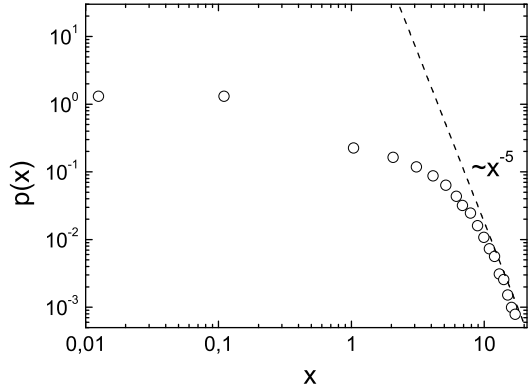}}
\caption{Probability density function for CTRW with power-law densities for waiting times and for resetting times under incomplete resetting at $t=1000$.  
The parameters are the same as in Fig. \ref{Gpdfc}: $\alpha=0.5, \beta=1.5$, $t_0=1$, $\tau_0=1000$.
The wing of the distribution now scales according to $p(x)\sim x^{-3-2\beta/\alpha+2/\alpha}$ corresponding to the domain 4 in Fig. \ref{fig:Domains}.\label{Gpdfi} } 
\end{center}
\end{figure}

The overview about intermediate power-law asymptotics of the PDFs is given in Fig. \ref{fig:Domains}.
The examples of such asymptotics as seen in numerical simulations, are given in Figs. \ref{Gpdfc} and \ref{Gpdfi}.

\section{Conclusions}
\label{sec:Concl}

We have studied subdiffusive continuous time random walks (CTRWs) with power-law resetting. We have considered the incomplete resetting, when the waiting period of CTRW is
unaffected by the resetting event and complete resetting, when the waiting period starts anew at the resetting event. We have shown that the behavior of MSD in CTRW under 
resetting is the same as for subdiffusive SBM under the same conditions which reflects the fact that SBM can serve as a mean field approximation for the CTRW for both cases. 
The PDF of displacements in CTRW under complete resetting is similar to such for the renewal SBM \cite{AnnaRenewal}. The fact that for both SBM and CTRW with complete resetting the forms of the MSD and PDF are similar is highly non-trivial (note that the free SBM is a Gaussian process, while the free CTRW possesses the PDF with a cusp and stretched-Gaussian tails). On the contrary, for the CTRW with incomplete resetting the behavior of the PDF of displacements differs considerably from the one for SBM \cite{AnnaNonrenewal} due to fluctuation effects connected with the distribution of the waiting time for the first jump of
CTRW after resetting.

\appendix

\section{Asymptotic properties of the integral, given by Eq.~(\ref{I1}).}
\label{sec:AppB}
At several places of our calculations (e.g., Eq.~(\ref{I1})) the integrals of the form
\begin{equation}
 I_1 (\alpha, \beta; z) = \int_0^1 y^\alpha(z + y)^{-\beta} dy
\end{equation}
with $\alpha > -1$ and different values of parameter $\beta$ appear, for which we are typically interested in their asymptotics when $z \to 0$. 

This integral is essentially given by a hypergeometric function, see Eq.(1.2.4.3) of \cite{BrPr}:
\begin{equation}
 I_1 (\alpha, \beta; z) = (1+\alpha)^{-1} z^{- \beta} \;_2 F_1 (1+\alpha,\beta;2+\alpha;- z^{-1}). \label{eq:IntI} 
\end{equation}
To get the asymptotic behavior of this function we can apply Pfaff transformation to transform the function of the argument $x=-z^{-1}$ which tends to infinity
to a function of the argument $\frac{x}{x-1} = \frac{-z^{-1}}{-z^{-1}-1} = \frac{1}{z+1}$ which tends to unity and use the special value of the hypergeometric function at
unity given by the Gauss theorem:
\begin{equation}
 \;_2 F_1 (a,b;c;1) = \frac{\Gamma(c) \Gamma(c-a-b)}{\Gamma(c-a) \Gamma(c-b)}
\end{equation}
which holds for 
\begin{equation}
 \Re c > \Re(a + b). 
 \label{eq:condition}
\end{equation}

There are two such Pfaff transformations:
\begin{equation}
\;_2 F_1 (a,b;c;x) = (1-x)^{-b} \;_2 F_1 \left(b,c-a;c;\frac{x}{x-1} \right)
 \label{eq:Pfaff1}
\end{equation}
and
\begin{equation}
\;_2 F_1 (a,b;c;x) = (1-x)^{-a} \;_2 F_1 \left(a,c-b;c;\frac{x}{x-1} \right),
 \label{eq:Pfaff2}
\end{equation}
with the first one being useful for $\alpha > \beta - 1$ (when the condition Eq.(\ref{eq:condition}) for the transformed of the hypergeometric function in Eq.(\ref{eq:IntI})
holds),
and the second one for the opposite case $\alpha < \beta -1$.
Applying the first transformation we get for $\alpha > \beta - 1$
\begin{equation}
 I_1(\alpha, \beta;z) \simeq z^0 \frac{1}{2+\alpha-\beta}
 \label{eq:AB1}
\end{equation}
i.e. tends to a positive constant. 

The second transformation which applies for $\alpha < \beta -1$ gives
\begin{eqnarray}
I_1(\alpha, \beta; z) &\simeq& z^{1+\alpha - \beta} \frac{1}{1+\alpha} \frac{\Gamma(2 + \alpha) \Gamma (\beta - \alpha - 1) }{\Gamma(1)\Gamma(\beta)} \nonumber \\
&=& z^{1+\alpha - \beta} \mathrm{B}(\alpha +1, \beta - \alpha -1),  \label{eq:AB2}
\end{eqnarray}
i.e., $I_1(\alpha, \beta; z) $ behaves as  $z^{1+\alpha - \beta}$.\\

\section{The Mainardi function.}

Let us start from our Eq.(\ref{eq:GFasy}):
\begin{equation}
 p(k,s) = \frac{s^{\alpha - 1}}{k^2 \tilde{K}_\alpha  + s^\alpha}
\end{equation}
and perform the inverse transforms. We first perform the inverse Fourier-transform by noting that
\begin{equation}
 \mathcal{F}^{-1} \frac{1}{k^2 + 1} = \frac{1}{2}e^{-|x|}.
\end{equation}
Therefore 
\begin{equation}
 p(x,s) = \frac{1}{2} \frac{s^{\frac{\alpha}{2} - 1}}{\sqrt{\tilde{K}_\alpha}} \exp \left(-\frac{|x|}{\sqrt{\tilde{K}_\alpha}} s^{\frac{\alpha}{2}} \right).
\end{equation}
Now we expand the exponential and perform term-per-term inverse Laplace transform, noting that
\begin{equation}
 \mathcal{L}^{-1} s^\gamma = \frac{1}{\Gamma(-\gamma)} t^{-\gamma - 1} 
\end{equation}
(for $\gamma$ not being a non-negative integer). Therefore we obtain
\begin{eqnarray*}
&& p(x,t) = \\
&& \frac{1}{2 \sqrt{\tilde{K}_\alpha}} \sum_{n=0}^\infty \left(- \frac{|x|}{\sqrt{\tilde{K}_\alpha}} \right)^n \frac{1}{n! \Gamma[- \frac{n \alpha}{2}  + (1 -\frac{\alpha}{2})]} t^{- \frac{(n+1) \alpha}{2}} = \\
&& \frac{1}{2 \sqrt{\tilde{K}_\alpha} t^{\alpha/2}} \sum_{n=0}^\infty \frac{1}{n! \Gamma[- \frac{n \alpha}{2}  + (1 -\frac{\alpha}{2})]} \left(- \frac{|x|}{\sqrt{\tilde{K}_\alpha} t^{\alpha/2}} \right)^n .
\end{eqnarray*}
The series in this asymptotic form represents a known special function  \cite{Mainardi,Mainardi2}:
\begin{equation}
p_{w,j}(x,t) = \frac{1}{2 \sqrt{\tilde{K}_\alpha} t^{\alpha / 2}} M_{\alpha / 2} \left(\frac{|x|}{ \sqrt{\tilde{K}_\alpha} t^{\alpha / 2} } \right), 
\end{equation}
with
\begin{equation}
 M_{\alpha / 2}(y) = \sum_{n=0}^\infty \frac{(-y)^n}{n! \Gamma[- \frac{n \alpha}{2}  + (1 -\frac{\alpha}{2})]} = \Phi_{-\alpha/2,1-\alpha/2}(y)
\end{equation}
being the Mainardi function, and $\Phi_{a,b}(y)$ being the Write function. 

The asymptotic behavior of $p(x,t)$ for small $|x|$ follows immediately from the series expansion of the Mainardi function. For all $\alpha < 1$ the function $p(x,t)$ (as a function
of $x$)
shows the cusp at zero which disappears only in the Gaussian case $\alpha = 1$ when all terms of odd orders disappear due to the divergence of the Gamma-functions
of whole non-positive arguments, and $M_{1/2}(y) =\pi^{-1/2} \exp(-y^2/4)$.

The asymptotic form of the Mainardi function \cite{Mainardi2} for $y$ large is
\begin{equation}
 M_{\alpha/2}(y) \simeq A y^a \exp(-b y^c)
\end{equation}
with
\begin{eqnarray*}
 A &=& [2 \pi (2 - \alpha) 2^{\alpha/(2-\alpha)}\alpha^{(2 - 2 \alpha)/(2-\alpha)}]^{-1/2} \\
 a &=& \frac{2 \alpha - 2}{2 (2-\alpha)} \\
 b &=& (2-\alpha) 2^{\alpha/(2-\alpha)} \alpha^{\alpha/(2-\alpha)} \\
 c &=& \frac{2}{2-\alpha}.
\end{eqnarray*}
Thus, for $|x|$ large compared to $\sqrt{\tilde{K}_\alpha} t^{\alpha / 2}$ the leading asymptotics of function $p(x,t)$ is 
\begin{equation}
p(x,t) \simeq C_1 \tilde{K}_\alpha^{-\frac{1}{2(2-\alpha)}} t^{-\frac{\alpha}{2(2-\alpha)}} |x|^\frac{2 \alpha - 2}{2 (2-\alpha)} \exp \left(C_2
\frac{|x|^{\frac{2}{2-\alpha}}}{\tilde{K}_\alpha^{\frac{1}{2-\alpha} } t^{\frac{\alpha}{2-\alpha} }}\right).
 \label{eq:Masympt}
\end{equation}
with constants $C_1$ and $C_2$ deriving from the previous expressions.

\end{document}